\documentclass{ws-ijmpa}
\usepackage[super,compress]{cite}
\usepackage{graphicx}
\usepackage{hyperref} 
\usepackage{CJKutf8} 
\usepackage[utf8]{inputenc}
\usepackage[T1]{fontenc}
\def\draftnote{}
\begin{document}
\markboth{Y. Shi}{Road of  Quantum Entanglement}
%
%

\title{The road of quantum entanglement: from Einstein to 2022 Nobel Prize in Physics 
}

\author{Yu Shi}
\address{Wilczek Quantum Center, Shanghai Institute for Advanced Studies, Shanghai 201315, China \\
University of Science and Technology of China, Hefei 230026, China\\
Department of Physics, Fudan University, Shanghai 200433, China\\yu\_shi@ustc.edu.cn 
}

\maketitle


\begin{abstract}
We explain the achievements that were awarded 2022 Nobel Prize in Physics, as well as the preceding and the later developments. The main notions and historic cornerstones of Bell inequalities, the related researches on quantum entanglement are reviewed, and the key physical ideas are emphasized.   Among the early work, C. S. Wu's contributions using polarization-entangled photons from electron-positron annihilation are introduced. \footnote{This review article is the English version of a Chinese article published in Chinese Journal of Nature {\bf 44} (6), 455-466 (2022) [
DOI: 10.3969/j.issn.0253-9608.2022.06.005], which was based on a talk the author gave at 2022 Fall Meeting of Chinese Physical Society (the video is available at https://www.koushare.com/video/details/173010). } 
 \end{abstract}

\ccode{PACS Nos: 03.67.Mn, 12.20.Fv, 01.65.+g}     

\vspace{1cm}

The 2022 Nobel Prize in Physics was awarded to Alain Aspect,
John F. Clauser, and Anton Zeilinger for their experiments using entangled photons
to verify the violation of Bell's inequality and for pioneering quantum
information science~\cite{1}. Alain Aspect is a 
professor at the Institut d'optique Graduate School,    Paris-Saclay University, and at the \'{E}cole Polytechnique in Paris, 
Clauser is  at Clauser and Associates, and Zeilinger is a professor at the University of Vienna. 
Their pioneering experiments made quantum entanglement a “powerful tool,” laying the foundation for a new era of quantum technology. 
This represents a new phase of the quantum revolution~\cite{2}, or the second quantum revolution~\cite{3}. 
 
Exactly 100 years before this Nobel Prize,  in 1922, Einstein received notification in Shanghai 
that he had been awarded the 1921 Nobel Prize in Physics (the prize had been
vacant in 1921)~\cite{4}. The Nobel Prize citation read:
``for his services to Theoretical Physics, and especially for his discovery of the law of the photoelectric effect''~\cite{5}  As an application of the photoelectric hypothesis,
the ``law of the photoelectric effect''   was part of his 1905 paper on light quanta
~\cite{6}.
Einstein once said, ``I have thought a hundred times as much about the quantum
problems as I have about general relativity theory"~\cite{7}. The quantum entanglement research that won the 2022 Nobel Prize in Physics
was pioneered by Einstein in 1935~\cite{8}. This article explains the achievements of the 2022 Nobel Prize in Physics
and outlines the main concepts and milestones of Bell's inequality research. We
emphasize that it was Einstein who pioneered the study of quantum entanglement, and this
great achievement is of primary importance.

\section{ Quantum Mechanics, Quantum States, and Photon Polarization}

Quantum entanglement is a concept in quantum mechanics. Quantum
mechanics is a  fundamental physical principle  governing microscopic particles such as electrons, photons, and atomic nuclei, 
while macroscopic objects typically follow classical physics.

Quantum mechanics originated in the early 20th century. Since the 1920s,
quantum mechanics has become the fundamental theoretical framework of all microscopic physics and has achieved tremendous success. The physical framework established before quantum mechanics
is referred to as classical physics. The mathematical tools of quantum mechanics
are no more complex than those of classical physics, but the conceptual framework of quantum mechanics
is fundamentally different. Heisenberg said, ``Anyone who is not shocked by quantum theory has not understood it''.  
In the history of human thought,
quantum mechanics has transformed the concept of reality and is one of the most significant revolutions.

The central concept of quantum mechanics is the quantum state ~\cite{9}. As the name suggests, 
a “quantum state” is a state described by quantum mechanics. It  is not
a physical quantity like mass or velocity, but rather a description similar to probability, or an extension of the concept of probability, 
from which a probability distribution can be calculated, yet it
contains more information than probability alone. Through quantum states,
many physical properties can be calculated. When measuring a property of a quantum system,
quantum state randomly changes with a certain probability (the original quantum state determines the magnitude of this probability)
to one of the quantum states that explicitly possesses this property.
Therefore, quantum states encompass various possibilities.

For example, the motion of a quantum particle in space is described by a spatial quantum state.  The particle  can be determined to be at a specific position, meaning
that its spatial quantum state represents it being at a specific position,
known as a position eigenstate. However, generally speaking, the quantum state describing spatial motion
is a superposition of different position eigenstates. That is,
when measuring its position, there is a certain probability of obtaining various positions,
thereby causing the quantum state to collapse into the corresponding position eigenstate. This
probability is equal to the square of the magnitude of the “position wave function.”

Mathematically, the spatial quantum state can be expressed as
\begin{equation} 
|\psi\rangle = \int \psi(x)  |x\rangle dx,
\end{equation}  
where $|x\rangle$  is the position eigenstate,  the integral over x represents the superposition of position eigenstates,  
 $\psi(x)$ is the position wave function representing the amplitude of $|x\rangle$, and $|\psi(x)|^2$  is the probability of measuring the position
to be $x$.

A quantum particle can also have a definite momentum, in which case the quantum state
is a momentum eigenstate. Generally, the same quantum state $|\psi\rangle$  describing spatial motion
can also be viewed as a superposition of different momentum eigenstates.  Therefore in measuring momentum. there is certain probability to obtain each possible value of the momentum and thus the quantum state collapses to the corresponding momentum eigenstate, with the probability being the square of magnitude of the  momentum wavefunction. 

Mathematically, this is expressed as
\begin{equation} 
|\psi\rangle =\int  \psi(x) |x\rangle dx =\int \phi(p) |p\rangle dp,  
\end{equation} 
where $|p\rangle$ is the momentum eigenstate, and integrating over p represents the superposition of momentum eigenstates, $ \phi(p)$ 
is the momentum wave function, and  $ |\phi(p)|^2$  is the probability of measuring momentum to be p.

The set of quantum states used to superpose the original quantum state is called the basis states, collectively referred to as the basis.  Which physical quantity is measured, an eigenstate of which the quantum state reduces to  after the measurement. 

For example, light has an internal property called polarization, which represents the direction of electric field  and is always located in a plane perpendicular to the direction of light propagation. In this plane, we can always arbitrarily establish a rectangular coordinate system, using unit-length arrows originating from the origin (called unit vectors) to represent the polarization, with the coordinates at the ends of the arrows representing the polarization direction.  Polarized sunglass only allows linearly polarized light whose polarization direction aligns with the lens's transmission axis to pass through.

Light is composed of light quanta (proposed as  energy quanta by Einstein in 1905,
 as mentioned earlier, and the work for which he received the Nobel Prize
was an application of this hypothesis. That a light quantum behaves like a particle with a momentum was proposed by Einstein in 1916). After 1926, light quanta have been also referred to as photons.
As a quantum particle, each photon has a polarization quantum state
corresponding to the macroscopic polarization of light. Any two mutually orthogonal
directions (such as those along two orthogonal coordinate axes) correspond to linearly polarized quantum states 
that are also mutually orthogonal, and any polarization quantum state can be superimposed in terms of them.

When a photon reaches a polarizing filter with a transmission direction along a certain direction, the photon either completely transmits or completely fails to transmit, and this occurs randomly. The probability of transmission is the square of the magnitude of the component of its original polarization quantum state corresponding to the transmission direction. After transmission, 
the photon's polarization quantum state collapses into the state corresponding to  the transmission direction.  If the photon does not transmit, the polarization quantum state  collapses into a state perpendicular to the transmission direction of the polarizer and is absorbed. The probability of  absorption is 1 minus that of transmission. Of course, the original polarization direction of the photon
may also happen to align with the transmission direction of the polarizer,
in which case the polarization quantum state remains unchanged.  

A polarization beam splitter (PBS) does not have the issue of photon absorption by the polarizer. There are two kinds of PBS's. A PBS of one kind utilizes the birefringence effect to decompose the incident polarized state into
two mutually orthogonal linear polarized states,  the corresponding light rays are called ordinary and extraordinary rays respectively, propagating in different directions because of different refractive indices. 

A PBS of the  other kind  is made from optically isotropic glass, and is made by cementing two half-cube prisms together, with the hypotenuse of one of them coated with a multilayer dielectric material.  
By using the interference effects of the coating layers,  the  state of the polarization parallel to or within  the incident plane (determined by the incident direction and  the normal of  the hypothenuse interface) transmits while the state of the   polarization orthogonal (senkrecht) to  the incident plane reflects, so that when a light is incident vertically, it is split into two components, the one  polarized within  the incident plane passes through the hypotenuse interface, 
and exits the PBS along the incident  direction;  the other component is polarized 
orthogonal to the incident direction, and is reflected, and propagates and exits  in the direction orthogonal to the incident direction (and also the polarization direction). 

When a single photon enters the PBS in perpendicular to the surface, it randomly
exits through one of two possible exits. If  it exits along the incident direction,
the polarization  must be  within the incident plane while orthogonal to the incident direction.    If it exits in the direction orthogonal to the incident direction, the polarization must be orthogonal to the incident plane and the incident direction.  

For example, we can use the horizontal and vertical polarization states $|\rightarrow \rangle$  and $|\uparrow \rangle$   as the basis states, or the $45^\circ$  and $135^\circ$  polarization states $|\nearrow \rangle$  and $|\nwarrow\rangle$    as the basis states, i.e.,
\begin{equation}
|\theta\rangle =\cos \theta |\rightarrow \rangle + \sin \theta |\uparrow \rangle=  \cos \theta' |\nearrow \rangle \rangle + \sin \theta' |\nwarrow \uparrow \rangle.
\end{equation}  
where $\theta$ is the angle relative to the horizontal direction, and $\theta'$ is the angle relative to the $45^\circ$ direction. We can also use circular polarization states  $|\pm\rangle$  as the basis vectors.
Relative to the direction of motion, $|+\rangle$  is right-handed, and $|-\rangle$ is left-handed, in other words,  their helicities are $1$ and $-1$, respectively.
We can obtain 
\begin{equation}
|\pm\rangle = \frac{1}{\sqrt{2}} (|\rightarrow\rangle \pm i |\uparrow\rangle ). 
\end{equation} 

When a large number of photons in the same polarization state arrive at a polarizer,
the proportion of photons that pass through the polarizer is equal to the probability of a single photon passing through.
Therefore, when a macroscopic electromagnetic wave arrives at a polarizer, part of it passes through along the transmission direction. For example, if the polarization state
is given by equation (3) and the transmission direction is horizontal, then the ratio of the transmitted
light intensity to the incident light intensity is $\cos^2\theta$. This is known as Malus' law. 
Correspondingly, the probability of a single photon transmitting is also $\cos^2\theta$.   
For convenience, this can also be referred to as Malus' law. The ratio of the energy of the original electromagnetic
magnetic wave energy (proportional to the ratio of the squares of the electric field magnitudes) is
the probability of a single photon transmitting.  The researches   recognized by this year's Nobel Prize in Physics 
all utilized  photon polarization.

\section{ Quantum entanglement and Einstein} 

Quantum entanglement is a property of the quantum state of a composite system. 
A composite system refers to a system composed of several subsystems. 
If at least one subsystem does not have an independent quantum state (strictly speaking, an independent quantum state refers to a quantum pure state, i.e., it does not 
contain a mixture in the classical probabilistic sense), then it is said that this 
subsystem is quantum entangled with the other subsystems, and this quantum 
state is called a quantum entangled state. In other words, quantum entanglement is a property of the quantum state of a whole composed of two 
or more subsystems, 
and this quantum state is called a quantum entangled state. In mathematics and quantum theory, the concept of quantum entanglement is clear. In fact, the vast
majority of quantum states are entangled.  

The earliest research on quantum entanglement was published in May 1935 by 
Albert Einstein and his two young colleagues Boris Podolsky and Nathan Rosen ~\cite{8}. They
did not use the term ``quantum entanglement",  but they discovered that quantum
entanglement was a property with special significance. In fact, quantum entanglement 
is also referred to as the Einstein-Podolsky-Rosen (EPR) correlation.

Einstein was a pioneer of quantum mechanics, but he was dissatisfied with the 
probabilistic framework of quantum mechanics and raised many objections. Gradually, 
his objections turned to focus  on the completeness of quantum mechanics. That is to say, 
that he believed certain elements of objective reality might not be described by quantum
 mechanics. His theoretical tool was exactly quantum entanglement. 
 
EPR considered a pair of particles originating from the same source,  moving in opposite directions 
while maintaining a total momentum of 0, at the same distance from the source,
i.e., with opposite displacements. Here I present a simplified form, setting the source's 
position to $0$, and representing the quantum states of the two particles as  
\begin{equation}
|\psi\rangle = \frac{1}{\sqrt{L}}\int |x\rangle |-x\rangle dx,  
\end{equation} 
where $L$ is some spatial range.   

There are many possibilities for the momenta  and positions of this pair of particles.  When measuring the position of the first particle, a random $x$ value is obtained with equal probability, and the position of the second particle can be immediately predicted to be $-x$. 

This quantum state  can also be written as 
\begin{equation}
|\psi\rangle = \frac{1}{h}\int |p\rangle |-p\rangle dp,  
\end{equation} 
which implies that if the momentum of the first particle is measured to be $p$,
then the momentum of the second particle can be immediately predicted to be $-p$. 

EPR paper states that after leaving the source, these two particles no longer interact with each other,  so measuring the first particle cannot affect the second particle, therefore  without disturbance, 
the position and momentum of the second particle can be deterministically predicted,
therefore  they are both intrinsic properties of the second particle, called   
elements of objective reality.

The above argument is based on the assumption that the description of quantum mechanics is complete,
leading to the conclusion that position and momentum are both objective elements of reality. However, in quantum mechanics,
the position and momentum operators do not commute, so it is impossible for both to have definite  values simultaneously and thus  both to be objective elements of reality, leading to a contradiction.
Based on this, EPR concludes that quantum mechanics is incomplete.

EPR also anticipated a counterargument, namely that someone might
argue that two or more physical quantities can only be said to be simultaneously objective elements of reality when they can be measured or predicted simultaneously, however, position  
and momentum are not measured simultaneously, a situation now referred to as counterfactual. 
But EPR deemed this unreasonable.

The title of the EPR paper is ``Can Quantum-Mechanical Description of Physical Reality Be Considered Complete?"   The abstract reads: ``In a complete theory there is an element corresponding to each element of reality. A sufficient condition for the
reality of a physical quantity is the possibility of predicting it with certainty, without disturbing the system. In quantum mechanics in the case of two physical quantities
described by non-commuting operators, the knowledge of one precludes the knowledge of the other. Then either (1) the description of reality given by the wave function in quantum mechanics is not complete or (2) these two quantities cannot have simultaneous reality. Consideration of the problem of making predictions concerning a system on the basis of measurements made on another system that
had previously interacted with it leads to the result that if (1) is false then (2) is also false. One is thus led to conclude that the description of reality as given by a wave function is not complete."

It is important to note that after measuring the first particle, only
the measurer of this particle can make predictions about the second particle; the controller of  the second 
particle is unaware of this, unless the measurer of the first particle
communicates the information to them, and this communication is constrained by relativity 
and other physical laws. This point is particularly important when discussing quantum information.

\section{Early follow-up work and polarization versions} 

EPR's work aroused Schr\"{o}dinger's great interest, and he had
several exchanges of letters with Einstein,  and published several papers in 1935. In one of them, ``Discussion on the Probabilistic Relations of Separated Systems,"  he wrote ~\cite{10}: ``When two systems, of which we know the states by their respective representatives, enter into temporary physical interaction due to known forces between them, and when after a time of mutual influence the systems separate again, then they can no longer be described in the same way as before, viz. by endowing each of them with a representative of its own. I would not call that one but rather the characteristic trait of quantum mechanics,  the one that enforces its entire departure from classical lines of thought. By the interaction the two representatives (or $\psi$-functions) have become entangled. "   

Schr\"{o}dinger also discussed the later famous ``Schr\"{o}dinger's cat"  paradox in another article, based on the quantum entanglement between nuclear decay and
the cat's life or death ~\cite{11}.

In October of the same year, Bohr also responded to EPR, with his complementarity principle~\cite{12}. The complementarity principle states that measuring two non-commuting physical quantities requires different measuring instruments. Bohr believed that because of the interaction between the object and the measuring instrument, the object's counteraction to the measuring instrument
cannot be controlled, classical causality must be abandoned; however, for the situation discussed in EPR paper, 
the uncertainty relation and the complementarity principle  remain applicable,
and the latter ensures that the quantum mechanical description satisfies all reasonable completeness requirements. 
Bohr specifically pointed out that in EPR's criterion for objective reality,
the phrase ``without disturbing the system"  is ambiguous. He said that
although the measurement of one particle does not have a mechanical
interaction, it still has an essential influence on the circumstances
under which the corresponding physical quantities are defined.

It is evident that Einstein sought to engage in a deeper discussion,
revealing the conflict between quantum entanglement and local realism (i.e., the coexistence of locality and realism). Locality refers to the idea that if the spatial distance between two events
exceeds the speed of light multiplied by the time interval—the so-called spacelike separation—then these two events cannot have a causal connection, as required by the theory of relativity. Realism means that 
observables are already determined before observation, independent of measurement.
EPR pointed out  that the physical quantities of the second particle,  predicted based on measurements of the first particle,
without disturbing the second particle, are objective elements of reality.
This conflicts  quantum mechanics.

Therefore, the study of quantum entanglement should be traced back to Einstein. Although later experiments refuted local realism, 
Einstein pioneered this field of research. We would like to say that the greatest contributor to the field of quantum entanglement research
is Einstein, as he once said elsewhere,
posing a question is often more important than solving it.

Bohr insisted that quantum mechanics is the entirety of the theory,
believing that objective reality is as it is, and disregarding the concepts of locality and realism.
His point was that when measuring the first particle, although there is no physical interaction acting on the second particle, the second particle is still
affected. This merely restates the rules of quantum mechanics without
addressing the EPR objections. Many physicists, based on Bohr's conclusion,
assumed the issue was resolved without delving into the details.

Einstein and others discussed position or momentum as continuous variables.
In 1951, David Bohm first used two spin-$\frac{1}{2}$  quantum states, which are simpler,  to discuss these issues ~\cite{13}, for example, 
\begin{equation} 
|\psi_{\pm}\rangle =
\frac{1}{\sqrt{2}} (|\uparrow\rangle |\downarrow\rangle \pm |\downarrow\rangle |\uparrow\rangle), \label{eq7} 
\end{equation}
which, together with other two entangled states 
\begin{equation} 
|\phi_{\pm}\rangle =
\frac{1}{\sqrt{2}} (|\uparrow\rangle |\uparrow\rangle \pm |\downarrow\rangle |\downarrow\rangle), \label{eq8} 
\end{equation}
are called Bell states, forming a basis for two spin-$\frac{1}{2}$.   Measurements on this basis are called Bell measurements. 

Spin-$\frac{1}{2}$ quantum states are similar to  the polarization states of photons, but since the spin quantum number is $\frac{1}{2}$,  $|\uparrow\rangle$ and  $ |\downarrow\rangle$ are orthogonal,   forming a basis.  For photon polarization, $|\leftarrow\rangle = - |\rightarrow\rangle$, they are physically the same quantum state with a phase difference. Instead,   $|\rightarrow\rangle$ and $|\uparrow\rangle$ are orthogonal and form a basis. 

For photon Bell states,  consider an   entangled state  as an example 
\begin{equation} 
\begin{array}{rl} 
\frac{1}{\sqrt{2}} (|\rightarrow\rangle |\rightarrow\rangle - |\downarrow\rangle |\downarrow\rangle)  &  =
-\frac{1}{\sqrt{2}} (|\nearrow\rangle |\nwarrow\rangle +   |\nwarrow\rangle |\nearrow\rangle)\\
&=\frac{1}{\sqrt{2}} (|+\rangle |+\rangle +   |-\rangle |-\rangle).
\end{array}
 \label{eq9} 
\end{equation}

If the polarization of the first photon is measured in order to see whether the result is  horizontal state   $|\rightarrow\rangle$ or the vertical state $ |\uparrow\rangle$,   then the result is obviously one of these two. If
the polarization of the first photon is indeed measured to be horizontal, then after this measurement, the polarization quantum state of the second
photon can be predicted to have collapsed to horizontal; if the polarization of the first
photon is measured to be vertical, then  after this measurement, the polarization quantum state of the second photon
can be predicted to have collapsed to vertical. If the polarization of the first photon
is measured in order to see whether the result is  $|\nearrow\rangle$ or  $|\nwarrow\rangle$, or whether the result is  $|+\rangle$ or  $|-\rangle$, the situation is similar.

Polarization is an internal property and is independent of spatial distance, so
two polarization-entangled photons can be very far apart. However, being very far apart 
means that during the separation, they are more susceptible to external disturbances, so entanglement is more easily disrupted.

We now use this entangled state to present the polarization version of the EPR argument.  Measure the polarization of the first photon as horizontal or vertical. If
it is measured to be horizontal (vertical), it can be clearly predicted 
that with 100\% probability,  the polarization of the other photon is also horizontal (vertical). Because the two photons
are separated by a spacelike distance (no physical signal transmission), the measurement of one photon does  not affect the second photon. Therefore, the horizontal-vertical polarization property of the second
photon is an    element reality (previously determined).
Similarly, measure the polarization of the first photon to be 45$^\circ$  or
135$^\circ$ . If it is measured to be 45$^\circ$  (135$^\circ$ ), it can be clearly predicted that  with 100\%  
probability,  the polarization of the other photon is 135$^\circ$  (45$^\circ$ ).
Because the two photons are separated by a spacelike distance (no physical signal transmission), 
measuring one photon does not affect the second photon. Therefore,
the second photon's 45$^\circ$ -135$^\circ$  polarization property is an  
element of reality  (previously determined).
In quantum mechanics, when the photon's polarization quantum state is horizontal
or vertical, both  45$^\circ$  and 135$^\circ$    are possible as the result of measuring the   45$^\circ$ -135$^\circ$    polarization property. Conversely, 
when the photon's polarization quantum state is  45$^\circ$  or 135$^\circ$,  both  horizontalness  and  verticalness  are possible as the result of measuring the   horizontal-vertical    polarization property.  This is because  the corresponding operators for  45$^\circ$ -135$^\circ$    and for horizontal-vertical   do not commute, and thus they cannot simultaneously have definite
values. Therefore, the   horizontal-vertical and 45$^\circ$ -135$^\circ$    properties  cannot
both be   elements of reality simultaneously . Thus, according to EPR's argument,
local realism contradicts the completeness of quantum mechanics. EPR asserts that local
realism is unshakable, so quantum mechanics is incomplete.

In 1957, Bohm and Aharonov pointed out ~\cite{14} that in 1950, Chien-Shiung Wu and her student I. Shaknov realized that 
photon polarization correlation (Bohm-Aharonov did not use the term “entanglement”) ~\cite{15}. Wu  and Shaknov measured the Compton scattering of the photon pairs produced by electron-positron  annihilation,
 accurately verifying quantum electrodynamics. The annihilation of a positron and   a electron produces two photons, whose polarizations  are always mutually orthogonal and are  scattered by electrons.   For different scattering angles,
measuring the two photons' motion directions in perpendicular and parallel cases
reveals the asymmetry of the probability distribution, i.e., the difference in probability 
between the two cases divided by the sum of them. The sensitivity  of the   $\gamma$  (photon)  detector of Wu   and Shaknov was 10 times that of previous
experiments, and they measured the asymmetry to be $2.04 \pm 
0.08$, which is very close
to the theoretical value of $2.00$.

Bohm and Aharonov also theoretically demonstrated that
non-entangled states cannot produce the experimental results of Wu   and Shaknov.  Although the focus of  Wu   and Shaknov was not on
quantum entanglement, the fact is that they were the first to experimentally realize
a clear and spatially separated quantum entangled state. They achieved
a photon polarization entangled state, which, using our current expression, is
$\frac{1}{\sqrt{2}} (|\rightarrow\rangle |\uparrow\rangle - |\uparrow\rangle |\rightarrow\rangle)$. 

\section{Local Realism and Bell's Inequality}

EPR argued that quantum mechanics is incomplete. This means that, in addition to
the quantum states in quantum mechanics, there are additional
variables of  the physical system that can describe the system's exact state. These additional 
variables are called hidden variables, and they represent what is known as realism. If
a theory that replaces quantum mechanics includes hidden variables, it is called
a hidden variable theory. If this theory also satisfies locality, it is
called a local hidden variable theory or local realism.

In 1931, von Neumann mathematically
proved that hidden variables do not exist ~\cite{16}. In the 1950s and 1960s, there were
some discussions about hidden variable theories, particularly Bohm's series of works.  In 1964 (published in 1966 due to editorial fault), Bell pointed out that
von Neumann's proof was not valid ~\cite{17}. 
In 1964, Bell further proposed that local realism is
inconsistent with quantum mechanics. He published an inequality that all local hidden variable theories 
should satisfy (published before his earlier work about von Neumann's proof)~\cite{18}. All subsequent
inequalities of this type are referred to as Bell's inequalities, which concern the correlation of measurement results between two subsystems, 
each of which is measured by a local observer. 
Calculating the correlation of various measurement results using local hidden variable theory yields results that satisfy Bell's inequalities.
However, in quantum mechanics, if these two subsystems are described using certain quantum entanglemed states, the results calculated according to quantum mechanics violate Bell's inequalities.  

Bell's discussion used spin-$\frac{1}{2}$ 
language, but it also applies 
to other similar systems, such as photon polarization. The observables
A and B of two particles are spin values divided by $\frac{\hbar}{2}$, taking values of either 1 or
-1, depending on the hidden variables and their respective measurement directions $\mathbf{a}$ and $\mathbf{b}$, so 
$A(\mathbf{a}, \lambda) = \pm 1$, $B(\mathbf{b},\lambda) = \pm 1$. According to local realism, their
correlation is 
$P(\mathbf{a},\mathbf{b}) = \int d\lambda \rho (\lambda)A(\mathbf{a},\lambda)B(\mathbf{b}, \lambda)$. For perfectly correlated A and B with  $A(\mathbf{a}, \lambda) = -B(\mathbf{a},\lambda)$, Bell proved that
\begin{equation}
1+P(\mathbf{b},\mathbf{c}) \geq  |P(\mathbf{a},\mathbf{b}) – P(\mathbf{a},\mathbf{c})|. 
\end{equation}
For spin-entangled state  $\frac{1}{\sqrt{2}} (|\uparrow\rangle |\downarrow\rangle - |\downarrow\rangle |\uparrow\rangle)$,  quantum mechanics gives $P(\mathbf{a},\mathbf{b}) = –\mathbf{a} \cdot \mathbf{b}$. By choosing appropriate values for a, b and c, we obtain a violation of the inequality. For polarization-entangled state 
$\frac{1}{\sqrt{2}} (|\rightarrow\rangle |\uparrow\rangle - |\uparrow\rangle |\rightarrow\rangle)$.   $P(a,b) = - \cos(\theta)$, where $\theta$  is the angle between a and b.

The foundational problems of quantum mechanics were once regarded as ``mere philosophy", 
but Bell's inequality proves that this is a theoretical and experimental
physics, transforming the originally metaphysical discussion into a
quantitatively determinable problem through experiments, and converting philosophical issues into quantitative scientific issues.

Experiments to test whether nature satisfies Bell's inequality are called
Bell tests. Conducting a Bell test requires the use of subsystems separated by a distance yet 
in a quantum entangled state, as well as rapid and efficient detection,
and the independent arrangement of each measuring  device in a manner
that cannot be predicted in advance.  All researches related to violations of Bell's inequality
(or Bell's theorem) have  been  based on Bell's pioneering work.

Experiments have determined that quantum mechanics is correct and that local realism is incorrect.
However, for a long time, there were logical flaws or
additional assumptions in the experimental determinations, which were only eliminated in recent years. The proposal and verification of Bell's inequality are closely related to the rise of quantum information science, including
concepts and experimental techniques. The achievements that won the 2022 Nobel Prize in Physics are  major contributions  to both of these areas.

\section{Bell-CHSH Inequality and Experiments}  

\subsection{ CHSH Inequalities}

Bell's original inequalities relied on  idealized assumptions,
such as perfect correlation, which could not be verified in actual experiments
and were therefore unsuitable for real experiments. In 1969, John Clauser,
Michael Horn, Abner Shimony,
and Richard Holt generalized Bell's inequality, commonly referred to as
the CHSH or Bell-CHSH inequality ~\cite{19}, making it more suitable for
real-world scenarios and testable in actual experiments.

Let's continue  our previous discussion of Bell's inequality. Consider
$A(a,\lambda)[B(b,\lambda)+ B(b',\lambda)]+ A(a',\lambda)]B(b,\lambda)-B(b',\lambda)]$, 
which must equal $\pm 2$, since either $B(b,\lambda)+B(b',\lambda)$ or $B(b,\lambda)–
B(b',\lambda)$ must equal $\pm 1$, while the other equals 0. Thus
we obtain $S = P(a, b) + P(a, b') + P(a', b) – P(a', b')$, which satisfies $–2 \leq S
\leq  2$.
This is the Bell-CHSH inequality. For Bell states,
S can reach $\pm2\sqrt{2}$. 

Therefore, as long as  local realism is valid, the Bell-CHSH inequality
holds and can be experimentally tested. Quantum mechanics violates
it, so whether quantum mechanics or local realism is correct depends on
which is consistent with experimental results.

Additionally, in 1989, Zeilinger collaborated with Daniel 
Greenberg and Michael Horne
discovered that a three-particle quantum entangled state possesses special
properties that conflict with local realism without requiring statistical averaging
~\cite{20}.

\subsection{Freedman-Clauser Experiment}

Clauser was a PhD student in molecular astrophysics at the time.
After earning his PhD in 1970, he joined the University of California, Berkeley, 
as a postdoctoral fellow under Charles Townes,
where he was allowed to independently study Bell's inequality. At Berkeley,  
in 1967, Carl Kocher, a student of Eugene Commins,
conducted his doctoral thesis research on
the time correlation of photon pairs from the same atomic source ~\cite{21}. 

In this system, cascade transitions produce entangled photon
pairs. An outer-shell electron of a calcium atom is excited from the ground state $4^{1}S_0$ to 
$6^1P_1$, then transitions to $6^1S_0$, and   to $4^1P_1$, emitting
a photon; then transitions back to the ground state $4^1S_0$, emitting another photon.  To maintain parity conservation with even parity, and angular momentum conservation as 0, the polarization state of the two photons must be  $\frac{1}{\sqrt{2}} (|+\rangle |+\rangle +   |-\rangle |-\rangle)$.   
In this entangled state, the coincidence rate of photons detected on both sides is $R(\phi)= \frac{1}{\sqrt{2}}\cos^2\phi$,    where $\phi$  is the angle between the directions of polarizations for the photon  detection on the two sides.   However, the  angles chosen by Kocker for detecting the photons
were 0° and 90°, which cannot be used to test the Bell
inequality.

Clauser and Eugene Cummins' doctoral student Freedman
modified the experimental setup and improved the efficiency of the polarizers.
 In this system, the CHSH inequality yields
 \begin{equation} 
 \left| \frac{R(22.5^\circ)}{R_0} - \frac{R(67.5^\circ)}{R_0} \right| - \frac{1}{4} \leq 0,
 \end{equation}
 where $R_0$  is the coincidence rate without polarizers.
 
In Clauser and Freedman's experiment, the left-hand side of the above equation
was $0.050\pm 0.008$, violating the Bell inequality, with a precision of 6
standard deviations ~\cite{22}.

This preliminary experimental attempt has flaws and limitations. Because
the efficiency of particle generation and detection is low, and measurements are pre-set, it is logically possible that hidden variables could selectively detect particles,
leading to a violation of the Bell inequality, moreover   
the locality requirement was not satisfied.

Locality is a key presumption of Bell's inequality. 
Measurements of two spatially  separated subsystems must be independent of each other, including 
the choice of which measurement to perform, such as position or momentum, transverse magnetic moment
or longitudinal magnetic moment (magnetic moment is proportional to spin), or the transmission direction of the polarizer. 
Therefore, the time difference between the two measurements must be sufficiently small 
so that no physical signal can travel from one side to the other within this time difference. Since
all signal speeds do not exceed the speed of light, experimentally, the time difference between the two measurements must be less than the distance divided by the speed of light.
The fixed setup of the Friedman-Clauser experiment does not satisfy the locality requirement.

\subsection{Aspect Experiment}

In 1981–1982, Alain Aspect and his collaborators
Philippe Grangier, Gerard Roger, and Jean Dalibard
conducted three experiments that observed violations of the Bell-CHSH
inequality with high precision, largely achieving locality.

In the first experiment ~\cite{23}, before the cascade process occurred,
two sets of lasers were used to directly excite electrons to the $6^1S_0$ state via two-photon absorption, 
which was much more effective than the previous method using $6^1P_1$.
In the second experiment ~\cite{24}, measurements were taken using a two-channel polarizer,
yielding good statistics and a significant violation of the Bell inequality,
with precision of several dozen standard deviations.

The third experiment was the most important ~\cite{25}. The distance from the calcium atom to the polarizer 
was 6 m, and the photons traveled in only 20 ns. Rotating the polarizer
during photon flight was not feasible. They employed a clever method designed by Aspect earlier 
~\cite{26}. A pair of photons passed through an acousto-optic switch before reaching a pair of polarizers,
directing them to one of two pairs of polarizers. The acousto-optic switch
switched every 10 ns. The CHSH inequality used
gives $-1 \leq S \leq 0$, while quantum mechanics gives $0.112$.
The experiment yielded $0.101 \pm  0.020$, consistent with quantum mechanics,
violating the inequality. The precision is $5$ standard deviations.

These experiments, along with many subsequent Bell test experiments,
conclude that quantum mechanics triumphs and local realism fails; however, these 
works still contained technical logical flaws, such as in detector
efficiency or locality. In Aspect's third experiment, the distance between the two
instruments was very short, and due to technical limitations, the changes in the measurement
devices were not random but periodic, thus failing to
close the locality loophole.

\subsection{   Zeilinger group closed the locality loophole in 1997 } 

In 1997,  Zeilinger group's experiment finally closed
the locality loophole ~\cite{27}. In their experiment, the devices analyzing entangled photon
pairs were separated by 400 m,  requiring 1,300
ns for light to travel. The entangled photon pairs were transmitted through optical fibers to polarizers. The direction of each photon's
polarization analysis device was rapidly and randomly changed, controlled by a random number
and timed using an atomic clock.
This experiment involved numerous technical improvements. It is worth noting
that the type-II parametric
down-conversion method was used to generate the entangled photon pairs. This is
a nonlinear optical process realized using $\beta$-boron borate ($\beta$-BBO) crystals.
The $\beta$-BBO crystals were first discovered and developed by Fujian Institute of Research on the Structure of Matter,  Chinese Academy of Sciences. In an earlier collaboration between Zeilinger's  group and Yanhua Shih, this method for generating entangled photon pairs was used ~\cite{28}. This method had been first proposed by Zeyu Ou and L. Mandel, as well as Yanhua Shih and C. O. Alley in the 1980s ~\cite{29-30}.

\subsection{Follow-up Work}

In experiments involving Bell inequalities, there has long been a “detection loophole.” This is because the entangled pairs of particles detected are only part of the originally generated entangled pairs, and the number of particles detected depends on the experimental setup.  Under the premise of fair sampling, the statistical analysis obtained from the experiment
can be used to test Bell's inequality. However, if
the detection efficiency is not high enough, fair sampling may not be achievable.
This is the detection loophole. To address the detection loophole and ensure fair
sampling, the following condition must be met: when a photon is measured on one side,
the probability of detecting a photon on the other side must be greater than 2/3 ~\cite{31}.
In 2001 and 2008, researchers closed the detection loophole in ion experiments ~\cite{32-33}.
In 2013, the Zeilinger group ~\cite{34} and the Kwiat group ~\cite{35} also closed the detection loophole in photon experiments.

In 2015, several experiments simultaneously closed both the locality loophole
and the detection loophole. The Zeilinger group ~\cite{36} and the Shalm group ~\cite{37} at the National Institute of Standards and
Technology (NIST) both used rapidly
switchable polarizers and high-efficiency photon detectors. The Hensen group at Delft
University of Technology used two pairs of electron-photon
pairs ~\cite{38} to measure two photons, causing the two electrons to become entangled. In 2017,
Weinfurter used entangled atoms separated by 398 m to simultaneously close
both loopholes ~\cite{39}.

Next, we introduce the “free choice loophole.” Bell's inequality
concerns the correlation between various measurement results of two subsystems,
involving several different settings of the measurement apparatus, such as the method of measurement. This is completely free in the derivation of Bell's inequality and
unrelated to hidden variables.

In Bell tests, it is necessary to freely and randomly select these different
settings. Even if the locality loophole and detection loophole are closed, in the experiment,
 it is the instrument that randomly selects the arrangement of the experimental apparatus. However, this is
not ideal, because what if the choices made by these instruments are themselves determined by
hidden variables? This is called the “free choice loophole.” Bell
proposed using human free choice to ensure the unpredictability of the experimental setup,
but the technology at the time could not achieve this.

On November 30, 2016, an experiment called “The Big Bell Test”
was conducted, addressing this “free choice loophole.” The choices made in the experiment
were provided by approximately 100,000 volunteers from around the world.
Over 12 hours, these volunteers participated in an online game called "the BIG
Bell Quest," generating 1,000 bits of data per second, totaling
97,347,490 bits of data. Volunteers were asked to
enter a certain number of random bits (0 or 1) within a specified time frame,
which were used as instructions for the choices made in the experiment. A machine learning algorithm
analyzed the entered bits and prompted volunteers to avoid predictable
but did not select the generated data.

Twelve laboratories across five continents conducted 13
Bell experiments within 12 hours. These experiments used the
randomly provided data from 100,000 volunteers to arrange
the measurement devices, with different experiments using
different datasets. The results of Bell tests across different
systems indicated that local realism was violated in these
systems, one of which was the photon polarization experiment
led by  Jianwei Pan.

On May 9, 2018, Nature magazine published the results of these 13 Bell experiments under the title “Challenging Local Realism with Human Choice” ~\cite{40-41},
showing that local realism is violated in systems involving photons, single atoms, atomic ensembles, and superconducting devices. This work represents another step forward in testing the foundational  theories of quantum mechanics.

Finally, since local realism conflicts with quantum mechanics,
where does the source of the contradiction lie—in locality or
realism? To investigate this question, Leggett considered a
“crypto-nonlocal realism”: as a form of non-locality, for
a given polarization direction, the measured quantity depends
both on the polarization direction of the measuring polarizer and
on the polarization direction of the other polarizer  However, the physical state
is the statistical average of various polarization directions, obeying local laws such as
Malus' law. For this, Leggett derived the Leggett inequality,
which is violated by quantum mechanics ~\cite{42}. Recently, we proposed a generalized Leggett inequality,
particularly applicable to entangled mesons in particle physics,
which is violated by both quantum mechanics and particle physics ~\cite{43}.

\section{Quantum entanglement is a resource in quantum information}

With the development of quantum physics and related technologies, particularly
research into the fundamental issues of quantum mechanics, quantum information science has gradually
emerged. Among these, the study of Bell's inequality and quantum entanglement has played
a significant role, demonstrating the importance of quantum entanglement. Quantum entanglement
has become a resource for quantum information processing ~\cite{9,44-46}. For example,
quantum entanglement can be used to achieve quantum teleportation.

Based on the principle of quantum superposition in quantum mechanics, a fundamental theorem in quantum information science is known as the “no-cloning theorem”:
it is impossible for a machine based on quantum mechanical unitary evolution to
clone an arbitrary  unknown quantum state ~\cite{47-48}. If
such a machine existed, it would be represented as a unitary  operator $U$, and the cloning process
would be $U |\psi\rangle |\phi\rangle|M\rangle = |\psi\rangle|\psi\rangle |M'\rangle$, 
where $|\psi\rangle$ is the state being cloned, $|\phi\rangle$ represents
the blank state before cloning,    $ |M\rangle$ and  $ |M'\rangle$,  as well as the following  $ |M''\rangle$ and  $ |M'''\rangle$  represent the machine's quantum states. Similarly, for another 
replicated state $|\psi'\rangle$, the replication process is $U |\psi'\rangle |\phi\rangle|M\rangle = |\psi'\rangle|\psi'\rangle |M'\rangle$. For $\alpha |\psi\rangle+\beta|\psi'\rangle$,  the replication process should be
$U (\alpha |\psi\rangle+\beta|\psi'\rangle) |\phi\rangle|M\rangle = (\alpha |\psi\rangle+\beta|\psi'\rangle) (\alpha |\psi\rangle+\beta|\psi'\rangle) 
 |M'''\rangle$, but 
the linear superposition principle of quantum mechanics implies that $U (\alpha |\psi\rangle+\beta|\psi'\rangle) |\phi\rangle|M\rangle = \alpha |\psi\rangle |\psi\rangle |M'\rangle +\beta|\psi'\rangle  |\psi'\rangle 
 |M''\rangle$,    which differs from the
expected cloning process. Therefore, no cloning  machine exists.
Thus, if an arbitrary quantum state is transferred from one carrier
to another through a certain process, the quantum state on the original
carrier must have changed. This is manifested in quantum teleportation.

In 1993, Bennett, Brassard, Cr\`{e}peau,
Jozsa, Peres, and Wootters proposed a quantum teleportation scheme,
using quantum entanglement and classical communication to transfer an unknown quantum state from the first
particle to a distant second particle ~\cite{49}. The third particle
is at the same location as the first particle but is entangled with the second particle in a Bell state, which can be represented as
$|\psi_+\rangle$  without loss of generality. Let the state of the first particle be denoted as $\gamma$. The quantum states of the three particles are (written in the order of first, third and second particles) 
\begin{equation} 
|\gamma\rangle|\psi_+\rangle =  \frac{1}{2}  |\psi_+\rangle |\gamma\rangle + \frac{1}{2} |\psi_-\rangle Z|\gamma\rangle + \frac{1}{2} |\phi_+\rangle X|\gamma\rangle +\frac{1}{2} |\phi_-\rangle XZ |\gamma\rangle,
\end{equation}
where  $X$ is the hermitian operator reversing $|\uparrow\rangle$   to $|\downarrow\rangle$ and vice versa,   $Z$ is the hermitian operator leaving $|\uparrow\rangle$ unchanged while multiplying  $|\downarrow\rangle$ with $-1$.  The inverse of each of these operators is itself. 

Alice controls the first and third particles and 
performs a Bell measurement (in the basis composed of the four Bell   states),   and
communicates, via some classical channel, the measurement results to Bob, who controls the second particle. 
Bob then performs the corresponding operation on the second particle.

After Alice performs the Bell measurement on the first and third particles, 
she knows that the state of the three particles has become which one of the four terms on the right-hand side of Eq.~12  above,
and she informs Bob of the result. Bob then performs the inverse  of the operation written before $|\gamma\rangle$ in the equation.
Thus, the final state of the second particle is always  $|\gamma\rangle$ .
If Alice obtains $|\psi_+\rangle$, Bob does nothing; if Alice obtains $|\psi_-\rangle$, 
Bob performs  Z; if Alice obtains $|\phi_+\rangle$,
Bob performs X; if Alice obtains$|\phi_-\rangle$,
Bob performs   ZX. 

The particle itself is not teleported; it is the quantum state that is teleported, the state of the  original carrier  (the first particle)  changes, becoming entangled with the third particle, And classical communication plays an important role. Thus, although
Alice and Bob do not know what  $|\gamma\rangle$  is, it is  teleported  from the first particle 
to the second particle, but without propagation through the space in between. Note that a key step is that  Alice 
notifies Bob  of the measurement result; otherwise, it is impossible to achieve quantum teleportation described here. The ingenuity
lies in the fact that neither Alice nor Bob knows the teleported state, and the particles themselves are not
teleported. 

Quantum entanglement and quantum teleportation cannot instantly transmit
information. If Alice does not notify Bob of the measurement result of her share of the entangled particles, the latter
cannot observe any changes in his share, and the measurement result is fully consistent with the quantum state before collapse (due to randomness). Therefore, there is no superluminal  transmission of signals, and quantum entanglement
does not violate relativity. Compliance with relativity is also reflected in
quantum teleportation, where Alice must inform Bob of the measurement results.

In fact, no signal transmission can exceed the speed of light.
In 1997, the Zeilinger group ~\cite{50} and the De Martini group ~\cite{51}
each experimentally demonstrated quantum teleportation.
As mentioned in the original theoretical paper on quantum teleportation,
quantum teleportation can be generalized as follows. Particles 1 and 2 are
in a Bell entangled state, and particles 3 and 4 are in another identical
Bell entangled state. Particles 2 and 3 are subjected to a Bell measurement together, resulting
in particles 1 and 4 being in an entangled state, even though they have not
interacted. This can be seen from the following equation:
\begin{equation} 
|\psi_+\rangle_{12} |\psi_+\rangle_{34} 
=\frac{1}{2} |\psi_+\rangle_{23} |\psi_+\rangle_{14} - 
\frac{1}{2} |\psi_-\rangle_{23} |\psi_-\rangle_{14}  
+ \frac{1}{2} |\phi_+\rangle_{23} |\phi_+\rangle_{14} 
- 
\frac{1}{2} |\phi_-\rangle_{23} |\phi_-\rangle_{14}. 
\end{equation}
A theoretical work Zeilinger participated referred to this as entanglement swapping and noted that it could  be used to detect the generation of entangled pairs ~\cite{52}.
In 1998, Zeilinger's group experimentally achieved entanglement swapping~\cite{53}.      Pan participated in the quantum teleportation
and entanglement swapping  experiments as a member of this  group.

One of the key goals of quantum technology is to achieve long-distance
quantum entanglement. One technical approach involves using optical fibers, but light
undergoes attenuation, so repeaters are required. However, quantum states cannot be
cloned, so this differs from classical repeaters.

One method involves using satellites, as in the free space above the atmosphere, 
light attenuation is minimal.    Pan's 
team in China used the Micius satellite launched in 2016 to implement this
scheme, achieving key distribution using the BB84 scheme
(Bennett and Brassard's 1984 scheme, which does not require quantum entanglement ~\cite{54})
between Xinglong near Beijing and Nanshan near Urumqi
(1,200 km apart) ~\cite{55}. Without using the satellite, but as a technical preparation for satellite-based applications,
they achieved quantum entanglement, quantum teleportation, and violation of the Bell-CHSH inequality
($S=2.51\pm 0.21$, with no loopholes for non-locality) over a distance of approximately 100 km near Qinghai Lake
~\cite{56}. Then, using the satellite, they distributed entangled photons to Delingha in Qinghai and Lijiang in Yunnan
Lijiang in Yunnan (1,203 km apart), observing two-photon entanglement
and violation of the Bell-CHSH inequality ($S=2.37\pm 0.09$, no
locality loophole)~\cite{57}. Later, in collaboration with the Zeilinger group, they achieved
key distribution between China and Austria (without quantum entanglement
)~\cite{58}. It is  also expected to achieve further breakthroughs using Satellites ~\cite{59}.

Another approach is the so-called quantum repeater, based on entanglement
swapping, enabling long-distance entanglement through multiple nodes. In addition to efficient 
entanglement swapping, good quantum storage is also required, as during multiple entanglement swapping processes on one side,
the other side must maintain the quantum state
unchanged. Combining these technologies could lead to the establishment of a global quantum network.  In 1991, Artur Ekert proposed a quantum key distribution scheme based on quantum entanglement states ~\cite{60}, known as the Ekert91 scheme. Suppose  Alice and Bob share entangled quantum bits (encoded in spin, photon polarization, or other carriers) in the state $|\psi_-\rangle$  originated from an independent source.
Then Alice and Bob each randomly measures their quantum bits in one of three directions $(\mathbf{a1}, \mathbf{a2}, \mathbf{a3}$  and two directions $(\mathbf{b1}, \mathbf{b2}, \mathbf{b3})$, respectively.  The $(\mathbf{a1}, \mathbf{a2}, \mathbf{a3})$ directions are
90$^\circ$ , 135$^\circ$ , and 180$^\circ$ , respectively, while the $(\mathbf{b1}, \mathbf{b2}$, $\mathbf{b3}$  directions are
135$^\circ$ , 180$^\circ$ , and 225$^\circ$ , respectively. The measurement results for the $(\mathbf{a1}$, $(\mathbf{a3}$, $(\mathbf{b1}$, and $(\mathbf{b3}$ directions
(which can be made public) are used to test the Bell-CHSH inequality. By
checking whether the Bell-CHSH  inequality is violated, it can be determined whether the channel is
secure and reliable and free from eavesdropping. Then,  the perfect anti-correlation measurement results in  $\mathbf{a2}$ (i.e. $\mathbf{b1}$) and $\mathbf{a3}$ (i.e. $\mathbf{b2}$)    are used to generate the key.
In 2006, the Zeilinger group achieved this scheme at a distance of 144 km ~\cite{61}.
They tested the CHSH inequality S, which was found to be $
2.508\pm 0.037$, indicating that the violation of the Bell inequality reached 13 standard deviations. In 2022, three groups used Bell tests without loopholes
to implement this scheme ~\cite{62-64}.

As a key generation scheme, it is also possible to measure the X or Z operators independently without testing the Bell inequality,
and the results should be anti-correlated ~\cite{65}.
Then, similar to the BB84 scheme, some results are used
for error rate analysis to examine whether there is  eavesdropping. If there is no eavesdropping,
a key can be generated. This is called the BBM92 scheme. In 2020,
the Micius satellite distributed entangled photons to Delingha and Nanshan (1,120 km apart),
implementing the Ekert91 and BBM92 schemes, and the violation
of the Bell-CHSH inequality was $S = 2.56 \pm  0.07$, reaching 8
standard deviations ~\cite{66}. In 2022, the Micius satellite will distribute entangled photon pairs
to Delingha and Lijiang (1,200 km apart), and then
achieve quantum teleportation between the two ground stations ~\cite{67}.
Quantum entanglement of multiple particles is the foundation
for achieving quantum computing and is also an important concept for understanding many-body quantum states~\cite{shi1,shi2,shi3}.

\section{Conclusion}

Einstein revealed the conflict between quantum mechanics and local realism,
Bell quantified it, and CHSH extended it to practical experiments.
To test Bell's inequality, experimental techniques have continuously improved.
Over time, this niche field has grown into a broader domain closely related to quantum manipulation 
and quantum information technology. Quantum entanglement
has become a powerful tool, laying the foundation for a new era of quantum technology,
and is the basis of what is called  the second quantum revolution  by some experts, or the  second  height of the quantum revolution  by myself, playing a crucial role in
quantum computing, quantum simulation, quantum communication, quantum metrology, and quantum sensing. 
As experimental work, the achievements recognized by the 2022 Nobel Prize in Physics 
have laid the foundation for the development of quantum technology. 

\section*{Acknowledgements}

This work was supported by National Natural Science Foundation of China (Grant No. T2241005).

\end{document}